\documentclass[aps,prb,twocolumn,superscriptaddress,showpacs]{revtex4}
\usepackage{graphicx}
\usepackage{amsmath,amssymb}
\usepackage{bm}
\begin{document}

\title{Energy gap opening in submonolayer lithium on graphene:
Local density functional and tight-binding calculations}

\author{M. Farjam}
\affiliation{Department of Nano-Science,
Computational Physical Sciences Laboratory,
Institute for Research in Fundamental Sciences (IPM), P.O.~Box 19395-5531,
Tehran, Iran}
\author{H. Rafii-Tabar}
\affiliation{Department of Nano-Science,
Computational Physical Sciences Laboratory,
Institute for Research in Fundamental Sciences (IPM), P.O.~Box 19395-5531,
Tehran, Iran}
\affiliation{Department of Medical Physics and Biomedical Engineering, and
Research Centre for Medical Nanotechnology and Tissue Engineering, Shahid
Beheshti University of Medical Sciences, Evin, Tehran 19839, Iran}

\date{\today}

\begin{abstract}
The adsorption of an alkali-metal submonolayer on graphene
occupying every third hexagon of the honeycomb lattice
in a commensurate $(\sqrt{3}\times\sqrt{3})R30^\circ$ arrangement
induces an energy gap in the spectrum of graphene.
To exemplify this type of band gap,
we present \textit{ab initio} density functional
theory calculations of the electronic band structure of C$_6$Li.
An examination of the lattice geometry of the compound system
shows the possibility
that the nearest-neighbor hopping amplitudes have alternating values
constructed in a Kekul\'e-type structure.
The band structure of the textured tight-binding model
is calculated and shown to reproduce
the expected band gap as well as other characteristic degeneracy
removals in the spectrum of graphene induced by lithium adsorption.
More generally we also deduce the possibility of energy gap opening in
periodic metal on graphene compounds C$_xM$ if $x$ is a multiple of 3.
\end{abstract}

\pacs{73.22.$-$f, 73.20.At, 81.05.Uw}

\maketitle

\section{Introduction}
The isolation of graphene, a honeycomb lattice of carbon atoms,
and the observation of the electric-field effect in the nanostructured samples,
deposited on oxidized silicon surface,
have renewed interest in the electronic properties of this
two-dimensional carbon material.
\cite{novoselov2004}
The breakthrough discovery made possible the experimental observation of
theoretically predicted exotic physics of graphene,
including an anomalous half-integer quantum Hall effect,
non-zero Berry's phase, and minimum conductivity.
\cite{novoselov2005b,zhang2005}

The exceptional properties of graphene, two-dimensional structure
and room-temperature high mobility,
make it an ideal material for carbon-based nanoelectronics envisaged in the
future nanotechnologies.
Various applications depend on the possibility of fabricating
a graphene-based field effect transistor (FET), including transistors
to be used in nanoscale electronic circuits,
light emitters and detectors, and ultra-sensitive chemical sensors and
biosensors.
\cite{avouris2007}
Being a zero-gap semiconductor, however, graphene
cannot be used directly as the conducting channel in FETs.
Discovering how to generate band gaps is therefore a question of fundamental
and applied significance.
Several possibilities exist.
Lateral confinement of the charge carriers in graphene nanoribbons (GNRs),
\cite{han2007}
and different gate voltages applied to the two layers of bilayer graphene
are promising methods of engineering band gaps in graphene.
\cite{mccann2006}
Recently, a band gap has been discovered in epitaxial graphene derived
from graphitization of silicon carbide.
\cite{zhou2007}
The effect of a substrate-induced band gap has also been investigated
experimentally in epitaxial graphene on nickel,
\cite{gruneis2008}
and theoretically in graphene on boron nitride.
\cite{giovannetti2007}

One of the routes toward tailoring the electronic properties of graphene is
through the adsorption of metals.
\cite{uchoa2008,giovannetti2008}
Alkali metals, in particular,
are donors of electrons and can be used for the purpose of doping
graphene to change its carrier concentration.
The related problem of adsorption of alkali metals on graphite,
which consists of
weakly linked layers of graphene, has been studied extensively in the past.
\cite{caragiu2005}
In a quite recent study, a band gap was experimentally observed,
and theoretically confirmed, in the surface
electronic structure of graphite, induced by the adsorption of sodium in
a $(5\times5)$ model.
\cite{pivetta2005}
The similar effect has also been reported for the $(4\times4)$ model.
\cite{rytkonen2007}
Due to the heavy doping, however,
the Fermi level is displaced into the graphene conduction $\pi^\ast$ band
so that the band gap lies in the occupied levels.
The gap-inducing mechanism in these cases
involves doping as well as interaction between the graphene
layers in the graphite surface.

In this paper,
we study the gap-inducing effect of alkali-metal adsorption on
single layer graphene, which represents a different problem than
graphite surface since the gap cannot be due to interlayer interactions.
For instance, first-principles calculations
of metal adsorption on graphene in the $(4\times4)$ model
\cite{chan2008}
do not show a gap at the Dirac point in the density of states (DOS).
We present the case of lithium on graphene compound C$_6$Li
in detail for several reasons.
This is a prototypical system in which the gap opening effect occurs,
it is a simple system to study theoretically, and it may be accessible
experimentally. The crucial property here is the structure of the system, so
that other alkali and alkaline-earth
metal compounds C$_6M$ can exhibit the same effect.
For instance, a similar band
gap has been calculated for calcium graphite intercalated compound,
\cite{calandra2005}
as well as for calcium on graphene, both as C$_6$Ca.
\cite{calandra2007}
Due to doping, lithium or another alkali metal
on graphene results in a metal and is not directly useful for semiconductor
applications.
It is, however, important to understand the mechanism responsible for the
gap opening effect.

An interesting question is which coverages of an alkali metal on graphene can
induce a band gap in its spectrum.
Our first-principles density functional theory (DFT) calculations
indicate the absence of a gap in
$(1\times1)$, $(2\times2)$, and $(4\times4)$
models, corresponding to C$_2$Li,
C$_8$Li, and C$_{32}$Li, respectively, but its presence in the
$(\sqrt{3}\times\sqrt{3})R30^\circ$ structure, which includes C$_3$Li
and C$_6$Li, as well as
in $(3\times3)$ coverage C$_{18}$Li.
Explaining the underlying reason is the subject of the following sections.

Our paper is organized as follows.
Section \ref{dft} presents the results of
DFT calculations. Section \ref{tbm} develops the tight-binding (TB) model.
It includes a description of the lattice
geometry, from which we clarify the gap mechanism within the TB model.
We conclude here that,
when lithium atoms are arranged above the hollow sites of
graphene honeycomb lattice
in a periodic commensurate structure to form C$_6$Li, the nearest-neighbor
hopping amplitudes acquire alternating values constructed in a Kekul\'e-type
structure.
This section also includes a discussion of the TB Hamiltonian and numerical
results for the Kekul\'e textured model with two hopping amplitudes.
Section \ref{con} contains a summary and conclusion.
The Appendix develops the solution of the TB model.

\section{DFT calculations} \label{dft}
Quantitative data for the electronic band structure of C$_6$Li were obtained
from DFT calculations,
using a plane wave basis set and pseudopotentials.
We used the \textsc{Quantum-ESPRESSO} code
and norm-conserving pseudopotentials,
\cite{espresso}
and performed the calculations with local-density approximation
(LDA) and Perdew-Zunger parametrization
\cite{perdew1981}
of the Ceperley-Alder correlation functional.
\cite{ceperley1980}
We used a lattice parameter of $a_0=2.46$~\AA\ for graphene, corresponding
to a C-C bond length of $1.42$~\AA.
Other theoretical calculations
\cite{chan2008}
have shown that alkali metals generally prefer to adsorb on the hollow sites of
graphene rather than the bridge or top sites.
We positioned the lithium atoms over the hollow sites of graphene in a
$(\sqrt{3}\times\sqrt{3})R30^\circ$ commensurate superstructure.
We remark, however, that lithium has not been shown to coat graphene or
graphite in this manner.
\cite{caragiu2005}
As a first approximation and in view of the strength of the graphene bonds,
initially we used the DFT code to only relax the distance of lithium
atoms from the graphene plane,
keeping the coordinates of carbon atoms fixed.
This yielded a distance of $1.79$~\AA\ between the lithium layer and graphene,
which is a somewhat smaller value, as expected for the LDA, than
that obtained in other
theoretical calculations based on generalized-gradient approximation (GGA).
\cite{rytkonen2007}
In our tight-binding model presented in the Sec.~\ref{tbm} we use the
data obtained for this structure.
However, since a change in the bond lengths has an important implication
for opening a band gap, we then
used the code to relax the positions of the carbon atoms as well as
the position of lithium atoms. This indeed resulted in a tiny Kekul\'e
distortion of C-C bond lengths, so that the near bonds to Li were contracted
by $2.8\times10^{-4}$~\AA, and the far bonds to Li were stretched by
$5.6\times10^{-4}$~\AA.
We then made band structure and DOS calculations for C$_6$Li.
For repeated images of the systems,
we used supercells, with $c=12$~\AA\ for graphene, and $c=15$~\AA\
for C$_6$Li.
A kinetic-energy cutoff of 70~Ry was needed for total-energy convergence.
Brillouin-zone (BZ) integrations were made with
a $6\times6\times2$ Monkhorst-Pack sampling of $k$-space,
\cite{monkhorst1976}
with Gaussian smearing of 0.05~Ry. For the calculation of the
DOS we used the tetrahedron method
\cite{blochl1994}
with a finer $k$-point mesh of $36\times36\times1$ grid.

\begin{figure}
\includegraphics[width=85mm]{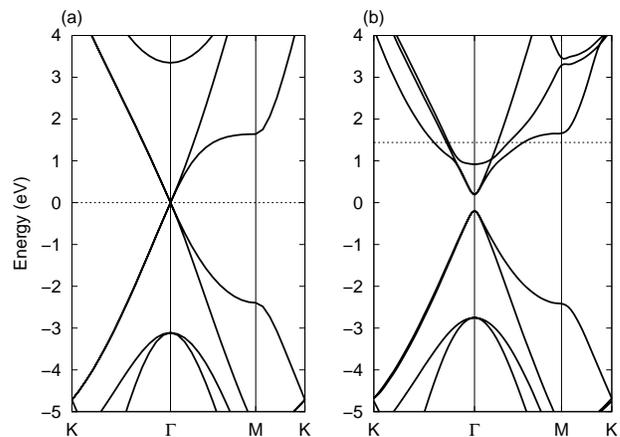}
\caption{(a) Band structure of C$_6\ast$, i.e., graphene described by
the unit cell of $(\sqrt{3}\times\sqrt{3})R30^\circ$ structure.
(b) Band structure of C$_6$Li. A band gap has opened at the neutrality point,
the degenerate bands along the $\Gamma$-$K$ direction are split, and a small
band gap has opened at $K$ near the bottom of the figure.
Horizontal dashed lines indicate the
Fermi levels of the respective systems.}
\label{fig1}
\end{figure}

The band structures of C$_6$Li, with unrelaxed carbon atom positions,
and C$_6\ast$, i.e., graphene described with the
unit cell of C$_6$Li, are compared in Fig.~\ref{fig1}.
We note the fourfold degeneracy of the bands at $\Gamma$, at
the Fermi level (Dirac or neutrality point),
in Fig.~\ref{fig1}(a),
which is due to the folding of $K$ and $K'$ points of
the graphene BZ onto $\Gamma$ in the BZ for
C$_6\ast$ structure.
We also note the linear dispersion of the energy bands
in the neighborhood of the Dirac point.
In Fig.~\ref{fig1}(b), we see a number of
changes that appear in the band structure of C$_6$Li.
Of interest for our study is the gap opening of $E_g=0.39$~eV
at the neutrality point.
For the C$_6$Li with fully relaxed positions, including those of carbon atoms,
the gap opening is $E_g=0.41$~eV, and otherwise the band structure is
quite similar to the unrelaxed case.
This means that the change in C-C bond lengths is responsible for only
about 5\% of the band gap.
We also note the lifting of other degeneracies,
at $K$ and along the $\Gamma$-$K$ direction.
Other changes that can be readily noted, as a result of lithium adsorption,
include the appearance of the parabolic-shaped band of lithium,
with its minimum located at $\Gamma$, at 0.92~eV above the neutrality point.
The lithium band weakly hybridizes with the graphene conduction bands,
resulting in some shifts and kinks in the bands of graphene.
There is a substantial charge transfer of $\agt0.2e$ per Li
atom from lithium layer to graphene, with every carbon atom receiving
the same charge of one-sixth of this value, i.e., $\agt0.03e$.
As a result, the Fermi level is raised by $1.5$~eV
relative to the neutrality point of graphene.
A rigid band model is clearly not a complete description,
but it can be used to describe the charge transfer,
which is another important aspect of this problem.

\section{Tight-binding model} \label{tbm}

\subsection{Lattice geometry}
\begin{figure}
\includegraphics[width=85mm]{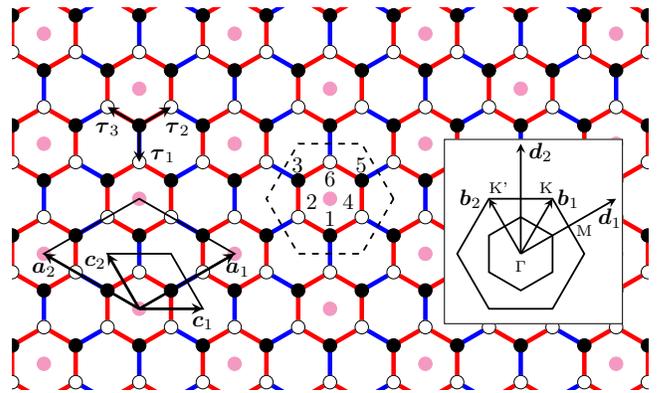}
\caption{(Color online) Lattice structure of lithium on graphene
C$_6$Li. Carbon atoms on sublattices $A$ and $B$ are depicted as filled
and open black circles.
Lithium atoms, in magenta,
occupy positions above the hollow sites.
The dashed hexagon is the Wigner-Seitz primitive unit cell of
C$_6$Li, containing six numbered C atoms and one Li atom.
Two types of bonds, two-thirds red and one-third blue, are distinguished
due to the presence of Li atoms.
On the lower left, the rhombus defined by $\mathbf{c}_1$ and $\mathbf{c}_2$
and enclosing two carbon atoms is the primitive
unit cell of graphene. The $30^\circ$ rotated
rhombus, defined by $\mathbf{a}_1$ and $\mathbf{a}_2$ and
having $\sqrt{3}$ times larger side, is the unit cell of C$_6$Li.
In the upper left part are shown the
vectors pointing from a carbon atom on sublattice $A$
to its three nearest neighbors on sublattice $B$.
The inset shows the reciprocal space.
The vectors $\mathbf{b}_{1,2}$ and $\mathbf{d}_{1,2}$ are sets
of reciprocal lattice vectors of C$_6$Li and graphene, respectively.
The hexagons in the inset are the Brillouin zones.
The region defined by $\Gamma{K}M$ is the irreducible wedge of graphene
Brillouin zone.}
\label{fig2}
\end{figure}

The band gap of C$_6$Li is a consequence of the special relationship
between its structure and that of the graphene substrate.
The Bravais lattice of C$_6$Li is hexagonal, as it is for graphene, but with a
$\sqrt{3}$ times larger lattice parameter and rotated by $30^\circ$, as shown
in Fig.~\ref{fig2}.
It is convenient to describe the base vectors of the hexagonal lattice
in a standard way, with the lattice parameter
used as unit of length, so for C$_6$Li we use the
coordinate system with its $x$-axis taken along $\mathbf{a}_1$
in Fig.~\ref{fig2},
\begin{equation} \label{base}
\mathbf{a}_1=(1,0)a, \qquad \mathbf{a}_2=
\left(-\frac{1}{2},\frac{\sqrt{3}}{2}\right)a,
\end{equation}
where $a=a_0\sqrt{3}$.
[Thus the same base vectors [Eq.~(\ref{base})] describe graphene,
i.e., $\mathbf{c}_{1}$ and $\mathbf{c}_{2}$
in Fig.~\ref{fig2}, with $a=a_0$ and with the $x$ axis
taken along $\mathbf{c}_{1}$.]
The corresponding reciprocal-lattice vectors of the hexagonal lattice are
\begin{equation} \label{reciprocal}
\mathbf{b}_1=\left(1,\frac{1}{\sqrt{3}}\right)\frac{2\pi}{a}, \qquad
\mathbf{b}_2=\left(0,\frac{2}{\sqrt{3}}\right)\frac{2\pi}{a}.
\end{equation}
We denote the reciprocal-lattice vectors of graphene, which are also given by
Eq.~(\ref{reciprocal}) relative to its own system, by $\mathbf{d}_{1,2}$.
The two sets of vectors $\mathbf{b}_{1,2}$ and $\mathbf{d}_{1,2}$
are related by a $30^\circ$ rotation and a $\sqrt{3}$ change in scale,
i.e., $|\mathbf{b}|=|\mathbf{d}|/\sqrt{3}$.

As shown in Fig.~\ref{fig2}, the $K$ points of graphene are located at
\begin{equation} \label{kk'}
K:\  \frac{1}{3}\mathbf{d}_1+\frac{1}{3}\mathbf{d}_2
\equiv\mathbf{b}_1, \quad
K':\ -\frac{1}{3}\mathbf{d}_1+\frac{2}{3}\mathbf{d}_2\equiv\mathbf{b}_2,
\end{equation}
i.e., the $K$ points of graphene coincide with reciprocal-lattice points
of C$_6\ast$. It also follows from Eq.~(\ref{kk'}) that the $K$ points of
graphene belong to the reciprocal lattice of the $(3\times3)$ structure.
Evidently, $(n\sqrt{3}\times{n}\sqrt{3})R30^\circ$ and $(3n\times3n)$
constructions, with $n$ a positive integer, will also share this property.
The existence of this kind of
relationship is the basic reason why a gap opens in C$_xM$ if $x$
is a multiple of 3, since in these structures the $K$ points
of the underlying graphene become coupled and the perturbation caused by
the foreign atoms removes the degeneracy.
The wave functions with wave vectors corresponding to $K$ and $K'$ of graphene
mix to form different standing waves.
One standing wave piles up electronic
charge in hexagonal cells occupied by Li, and
depletes it from the hexagonal cells devoid of Li,
and the other standing wave does the opposite.
(The commensurate structure of C$_6$Li breaks a $Z_3$ symmetry.)
These standing wave states then experience different potentials which
qualitatively explains the origin of the gap.

Each carbon atom of sublattice $A$ is connected to three carbon atoms of
sublattice $B$ by the $\bm{\tau}$ vectors, as shown in Fig.~\ref{fig2}.
These vectors in the coordinate system of C$_6$Li, needed in our
tight-binding calculations, are given by
\begin{equation} \label{tau}
\bm{\tau}_{1,3}=\left(-\frac{1}{6},\mp\frac{1}{2\sqrt{3}}\right)a,
\qquad
\bm{\tau}_2=\left(\frac{1}{3},0\right)a.
\end{equation}

We now examine the system from the viewpoint of the tight-binding model.
The presence of the lithium atoms, in C$_6$Li,
above the hollow hexagonal sites can be
imagined to have two effects, within a simple tight-binding model: (a) a
contribution (Hartree potential)
to the on-site energies of the C atoms, and (b) a possible change in
the hopping amplitudes depicted as C-C bonds in Fig.~\ref{fig2}.
From an inspection of the Wigner-Seitz cell in Fig.~\ref{fig2},
the C atoms are seen to occupy equivalent positions with respect to the metal
atoms since each is at a vertex shared by one metal-filled and two empty
hexagons of the honeycomb lattice.
Therefore all C atoms receive the same charge transfer from Li atoms,
as we have also seen in our DFT calculations, and
the symmetry breaking cannot be attributed to on-site energies.
This reduces the number of parameters and makes a TB model
description much simpler for C$_6$Li than for more dispersed compounds.
The bonds, however, occupy two kinds of positions
in a Kekul\'e construction, two-thirds of them (colored red)
between a filled and an empty hexagon, and one-third (colored blue)
between two empty hexagons.
We conclude that there are two different hopping amplitudes
corresponding to the two kinds of bond positions.
Our DFT calculations showed that changes in bond-length account for
$\simeq5$\% of the energy gap.
The modulation in hopping amplitudes, which are matrix elements
of the Hamiltonian between carbon $2p_z$ orbitals, is
therefore caused mainly by different potential energies and electron
concentrations in the regions of red and blue bonds,
with a small contribution due to bond-length distortion.
Of course, both variations in potential energy and the distortion of bond
lengths are caused by the presence of lithium ions, and their effects
on the band gap add with each other.

\subsection{TB Hamiltonian}
The $\pi$ bands of graphene are described well by a Hamiltonian of spinless
electrons hopping on the honeycomb lattice of graphene,
\begin{equation} \label{hamiltonian}
H=\sum_i \epsilon_ic_{i}^\dag c_{i}-\sum_{ij}t_{ij} c_{i}^\dag c_{j}.
\end{equation}
Here $c_{i}\ (c_{i}^\dag)$ annihilates (creates) an electron at site $i$,
$\epsilon_i$ are on-site energies, which are all equal and therefore
can be set to zero, and $t_{ij}$ are the hopping amplitudes.

The nearest-neighbor TB model of graphene has a gapless band structure, but a
Kekul\'e modulation of the hopping amplitudes couples the $K$ and $K'$ points,
leading to a mixing of degenerate states and opening of an energy band gap.
\cite{hou2007}
Following our discussion above, we allow
the hopping amplitudes $t_{ij}$ to have two values,
$t_1$ for two-thirds of
the bonds on the hexagonal cells beneath a metal atom, and
$t_2$ for the remaining one-third of the bonds.
Here we are interested in describing the gap, and therefore we exclude the
metal atoms from our TB model, except for their effect on modulating the
hopping amplitudes. They can be included in a more general treatment,
\cite{ishida1992}
but this is not needed for describing the gap of C$_6$Li.

\begin{figure}
\begin{center}
\includegraphics[width=85mm]{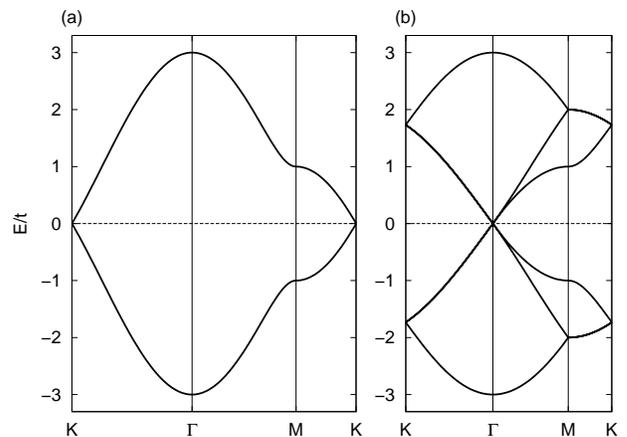}
\caption{Band structures of graphene in the Brillouin zones of (a)
the $(1\times1)$ structure, and (b)
the $(\sqrt{3}\times\sqrt{3})R30^\circ$ structure.}
\label{fig3}
\end{center}
\end{figure}

For clean graphene $t_1=t_2=t$.
There are two $\pi$ bands corresponding to the two-atom basis,
and the band energy can be written as
\begin{equation} \label{eband}
E_{\pm}(\mathbf{k})=\pm t\left|\sum_{i=1}^3
e^{i\mathbf{k}\bm{\cdot}\bm{\tau_i}}\right|,
\end{equation}
where the vectors $\bm{\tau_i}$ are defined in Fig.~\ref{fig2}.
When we use the unit cell of the
$(\sqrt{3}\times\sqrt{3})R30^\circ$ superlattice to describe graphene,
we will have six $\pi$ bands since there are six carbon atoms within this unit
cell. The additional bands are derived from Eq.~(\ref{eband}) as
$E_{\pm}(\mathbf{k}-\mathbf{b}_1)$ and
$E_{\pm}(\mathbf{k}-\mathbf{b}_2)$, with the
vectors $\mathbf{b}_{1,2}$ given by Eq.~(\ref{reciprocal}),
which result in the folding of $K$ and $K'$ points onto the
$\Gamma$ point of BZ, because of relation (\ref{kk'}).
The energy bands from these analytic formulas are plotted in
Fig.~\ref{fig3}. It is seen that some degeneracies are present at the new $K$
points [Fig.~\ref{fig3}(b)] as well as along $\Gamma{K}$ and $MK$ lines.
The linear dispersion near the Dirac point
is shown in the Appendix to be $E_{\pm}(\mathbf{k})=\pm\hbar{v_F}|\mathbf{k}|$,
where, $v_F=\sqrt{3}a_0t/2\hbar$ is the Fermi velocity, $t$ is the
nearest-neighbor TB hopping parameter,
and $a_0$ is the graphene lattice parameter.

\subsection{Numerical results}
We must verify that
the band gap of C$_6$Li can be described by a textured TB model
with two different hopping amplitudes, $t_1\ne{t_2}$.
In this case, we obtain the band structure by numerical diagonalization of the
Hamiltonian [Eq.~(\ref{hamiltonian})] in momentum space.
The method is described in the Appendix, where we
also derive an analytic expression
for the energy dispersion in the neighborhood of the gap as
\begin{equation} \label{gap}
E_\pm=\pm\left(\hbar^2{v_F}^2\mathbf{k}^2+\frac{E_g^2}{4}\right)^{1/2},
\end{equation}
where
\begin{equation} \label{tbgap}
E_g=2|t_2-t_1|
\end{equation}
for the textured TB model.
The Fermi velocity is given in terms of the TB parameter as before,
except that now an average value,
\begin{equation} \label{tbt}
t=\frac{2t_1+t_2}{3},
\end{equation}
must be used.

From the slope of the linear dispersion near the neutrality point of
Fig.~\ref{fig1}(a), we can find the Fermi velocity and then the TB parameter
$t=2.72$~eV, in good agreement with the commonly used range for this parameter.
\cite{reich2002}
From Fig.~\ref{fig2}(b), we can find
the renormalized Fermi velocity $v_F$ and thus the hopping amplitude $t$
by fitting Eq.~(\ref{gap}) to the band-structure data in
the vicinity of the gap.
We found that the value of $t$ is slightly larger for the upper branch than
that for the lower branch, by $\sim0.1$~eV, but that the change in the
average value is much smaller than this amount.
In the following, we neglect the renormalization of the
Fermi velocity and the average hopping amplitude.

\begin{figure}
\begin{center}
\includegraphics[width=85mm]{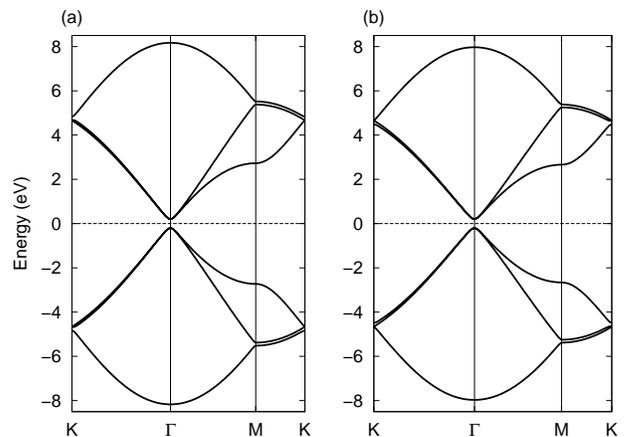}
\caption{Band structures of graphene from the TB model with two hopping
amplitudes:
(a) $t_1=2.79$~eV, $t_2=2.59$~eV, and (b) $t_1=2.59$~eV, $t_2=2.79$~eV.
Note that the three-fold degeneracies at $K$, e.g.,
the lower left one, are lifted
so that in (a) we have first a non-degenerate level
and above it a doubly degenerate level, and in (b) we have just the
opposite order of the degenerate and nondegenerate levels.}
\label{fig4}
\end{center}
\end{figure}

Given the values of $t=2.72$~eV and $E_g=0.39$~eV, from first-principles
calculations, we can determine $t_1$ and
$t_2$ from Eqs.~(\ref{tbgap}) and (\ref{tbt}). Thus we find
$t_1=2.79$~eV and $t_2=2.59$~eV.
Interchanging $t_1$ and $t_2$ is also a solution since $E_g$ depends
on the absolute value of their difference. To distinguish between the two, we
plot the band structures corresponding to both results in Fig.~\ref{fig4}.
The band gaps are identical in Figs.~\ref{fig4}(a) and \ref{fig4}(b), but a
characteristic difference can be seen in the way the threefold degeneracy is
lifted at $K$.
Comparing with Fig.~\ref{fig1}(b),
it is seen that Fig.~\ref{fig4}(a) has the same qualitative feature,
i.e., the same shape of band gap, and the lifting of degeneracies along
$\Gamma{K}$ and at the $K$ point.
This shows the validity of the Kekul\'e textured model for describing
lithium adsorbed graphene,
as well as showing that for C$_6$Li, $t_1>t_2$.
In fact, Fig.~\ref{fig4}(b) corresponds to our
DFT calculations of the band structure of C$_3$Li (not shown),
which has the same lattice structure as C$_6$Li, and is its complement.

\begin{figure}
\begin{center}
\includegraphics[width=85mm]{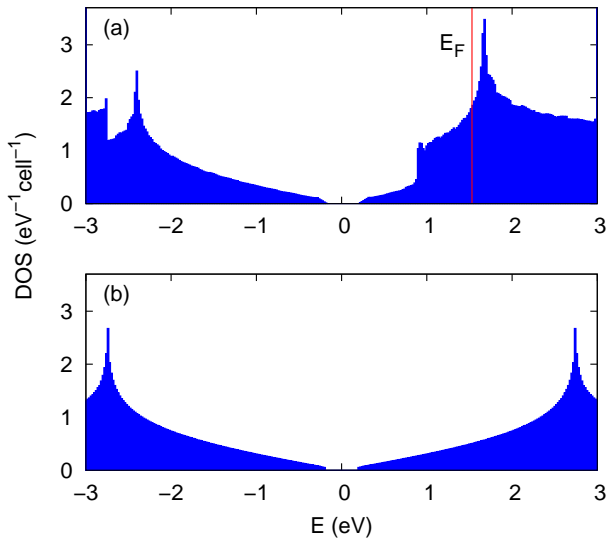}
\caption{(Color online)
(a) DOS of C$_6$Li calculated by DFT code. The vertical
red line indicates the position of the Fermi level. (b) The
corresponding DOS for graphene calculated within the tight-binding model.
The position of Van Hove singularities are displaced relative to the DOS of (a)
and the extra densities corresponding to Li atoms are not included in (b).}
\label{fig5}
\end{center}
\end{figure}

A useful quantity that can be calculated from energy bands obtained is
the density of states per unit cell given by
\begin{equation}
g(E)=2A\sum_{n=1}^6\int_\text{BZ}\frac{d\mathbf{k}}{(2\pi)^2}
\,\delta[E-E_n(\mathbf{k})],
\end{equation}
where $A$ is the area of the C$_6\ast$ unit cell.
For the TB DOS we used a $60\times60$ $k$-point mesh and the triangle method,
\cite{lee2002}
which is the two-dimensional version of the tetrahedron method
\cite{blochl1994}
that we also used for the \textit{ab initio} calculations.
The density of states calculated by first-principles DFT and
our TB model compare favorably in the vicinity of the gap,
as shown in Fig.~\ref{fig5}.
The extra densities of states in the energy range above the gap
in Fig.~\ref{fig5}(a)
are due to
the band of the lithium atoms, which are not included in our
tight-binding model.
The other main difference between the two results is the
positions of the Van Hove singularities which are closer to the gap in first
principles calculations.

\section{Summary} \label{con}
We have studied the opening of a band gap induced in single layer graphene
by the adsorption of a submonolayer of an alkali metal.
A band gap induced by the adsorption of alkali metals
was previously observed in graphite, and its origin was
explained to be due to charge transfer to substrate and interlayer interaction
of graphene layers.
In single layer graphene, where interlayer interaction does not exist,
the origin of an alkali-metal induced band gap
is in the coupling of $K$ and $K'$ points
which is only possible for appropriate superstructures.
The required property exists in the superstructures
$(3n\times3n)$ and $(n\sqrt{3}\times{n}\sqrt{3})R30^\circ$, with $n$ as a
positive integer.
For other coverages on graphene we expect the bands to remain
degenerate at the $K$ points of graphene, resulting in a gapless spectrum.
A tight-binding model with alternating hopping amplitudes forming the
Kekul\'e construction was shown to describe the characteristic features of the
band structure of C$_6$Li. From our DFT calculation, we concluded that the
modulation of the hopping amplitudes is mainly due to variations in charge
density in the graphene plane due to the attraction caused by the Li ions.
A tiny distortion of bond lengths additively contributes
$\sim5$\% to the band gap and, thus, enhances the modulation in
hopping amplitudes.
In conclusion, we showed that the band gap opening in the lithium on graphene
compound C$_6$Li is consistent with the nearest-neighbor
tight-binding model of graphene with Kekul\'e modulated hopping amplitudes.

\begin{acknowledgments}
M.~F. thanks S.~Farjam for graphical assistance and A.~Saffarzadeh for useful
discussions. First
principles computations were performed with the resources of the Computational
Nanotechnology Supercomputer Centre at the Institute for Research in
Fundamental Sciences (IPM).
The financial support of the Iranian Nanotechnology Initiative (INI)
is gratefully acknowledged.
\end{acknowledgments}

\appendix*
\section{Solution of TB model} \label{app}
In this appendix, we summarize our method of solving the
Kekul\'e textured nearest-neighbor tight-binding model
[Eq.~\ref{hamiltonian}].
First, we expand the eigenfunctions of our general periodic system,
characterized by crystal momentum $\mathbf{k}$,
in terms of the $2p_z$ orbitals, $|nj\rangle$, localized on the $j$th atom in
the $n$th unit cell, as
\begin{equation}
|\mathbf{k}\rangle=\frac{1}{\sqrt{N}}\sum_{n=1}^N \sum_{j=1}^M A_j
e^{i\mathbf{k}\bm{\cdot}\mathbf{r}_{nj}} |nj\rangle.
\end{equation}
Here $N$ is the number of unit cells, and $M$ is the number of carbon
atoms in the basis,
$\mathbf{r}_{nj}$ are position vectors of the atoms, and the amplitudes
$A_j$ form the components of the eigenvectors
to be determined by diagonalizing the Hamiltonian.
For graphene $M=2$, and the Hamiltonian matrix in momentum space is
given by
\begin{equation}
H(\mathbf{k})=-t
\begin{pmatrix}
0 & \sum_{i=1}^3e^{i\mathbf{k}\bm{\cdot}\bm{\tau_i}} \\
\sum_{i=1}^3e^{-i\mathbf{k}\bm{\cdot}\bm{\tau_i}} & 0
\end{pmatrix},
\end{equation}
which is readily diagonalized to give Eq.~(\ref{eband}).

For the textured model, $M=6$.
Referring to Fig.~\ref{fig2} for numbering of the basis atoms, and the
$\bm{\tau}_i$ vectors, we can write the $6\times6$ Hamiltonian matrix.
For example, to obtain the first row we observe that
atom 1 has nearest neighbors 2, 4, and 6 (due to periodicity), and
is connected to them by $\bm{\tau}_3$, $\bm{\tau}_2$, and
$\bm{\tau}_1$, respectively, with hopping amplitudes $t_1$, $t_1$, and
$t_2$. Thus we find
\begin{widetext}
\begin{equation} \label{hmat}
H(\mathbf{k})=-\begin{pmatrix}
0 & t_1e^{i\mathbf{k}\bm{\cdot}\bm{\tau}_3} &
0 & t_1e^{i\mathbf{k}\bm{\cdot}\bm{\tau}_2} &
0 & t_2e^{i\mathbf{k}\bm{\cdot}\bm{\tau}_1}\\
t_1e^{-i\mathbf{k}\bm{\cdot}\bm{\tau}_3} & 0 &
t_1e^{-i\mathbf{k}\bm{\cdot}\bm{\tau}_1} & 0 &
t_2e^{-i\mathbf{k}\bm{\cdot}\bm{\tau}_2} & 0 \\
0 & t_1e^{i\mathbf{k}\bm{\cdot}\bm{\tau}_1} &
0 & t_2e^{i\mathbf{k}\bm{\cdot}\bm{\tau}_3} &
0 & t_1e^{i\mathbf{k}\bm{\cdot}\bm{\tau}_2}\\
t_1e^{-i\mathbf{k}\bm{\cdot}\bm{\tau}_2} & 0 &
t_2e^{-i\mathbf{k}\bm{\cdot}\bm{\tau}_3} & 0 &
t_1e^{-i\mathbf{k}\bm{\cdot}\bm{\tau}_1} & 0 \\
0 & t_2e^{i\mathbf{k}\bm{\cdot}\bm{\tau}_2} &
0 & t_1e^{i\mathbf{k}\bm{\cdot}\bm{\tau}_1} &
0 & t_1e^{i\mathbf{k}\bm{\cdot}\bm{\tau}_3}\\
t_2e^{-i\mathbf{k}\bm{\cdot}\bm{\tau}_1} & 0 &
t_1e^{-i\mathbf{k}\bm{\cdot}\bm{\tau}_2} & 0 &
t_1e^{-i\mathbf{k}\bm{\cdot}\bm{\tau}_3} & 0
\end{pmatrix}.
\end{equation}
For general $\mathbf{k}\in\text{BZ}$, we diagonalize Eq.~(\ref{hmat})
numerically, but we obtain an analytic expression for the neighborhood
of the gap using perturbation theory.

At $\mathbf{k}=0$, the unperturbed Hamiltonian,
case of $t_1=t_2=t$, has four of its eigenvalues equal to $0$,
with the other two being $\pm3t$. The corresponding eigenvectors are easily
written down by noting that they are related to the $K$, $K'$, and $\Gamma$
points of graphene described by its primitive unit cell,
\begin{equation} \label{basis}
\frac{1}{\sqrt{3}}
\begin{pmatrix}
1 \\ 0 \\ \omega  \\ 0 \\ \omega^\ast \\ 0
\end{pmatrix}, \qquad
\frac{1}{\sqrt{3}}
\begin{pmatrix}
0 \\ \omega \\ 0 \\ \omega^\ast \\ 0 \\ 1
\end{pmatrix}, \qquad
\frac{1}{\sqrt{3}}
\begin{pmatrix}
0 \\ \omega^\ast \\ 0 \\ \omega \\ 0 \\ 1
\end{pmatrix}, \qquad
\frac{1}{\sqrt{3}}
\begin{pmatrix}
1 \\ 0 \\ \omega^\ast \\ 0 \\ \omega \\ 0
\end{pmatrix}, \qquad
\frac{1}{\sqrt{6}}
\begin{pmatrix}
1 \\ \pm1 \\ 1 \\ \pm1 \\ 1 \\ \pm1
\end{pmatrix},
\end{equation}
where $\omega=e^{i2\pi/3}$.

Next we calculate the Hamiltonian matrix in the fourfold
degenerate subspace of the zero eigenvalue, spanned by the first four vectors
of Eq.~(\ref{basis}), and expand the result near $\mathbf{k}=0$ to obtain
(cf. Ref.~23)
\begin{equation} \label{hgap}
\mathcal{H}(\mathbf{k})=
\begin{pmatrix}
0 & \hbar v_F(k_x-ik_y) & t_2-t_1 & 0 \\
\hbar v_F(k_x+ik_y) & 0 & 0 & t_2-t_1 \\
t_2-t_1 & 0 & 0 & -\hbar v_F(k_x-ik_y) \\
0 & t_2-t_1 & -\hbar v_F(k_x+ik_y) & 0
\end{pmatrix},
\end{equation}
where $v_F=\sqrt{3}a_0t/2\hbar$ as for unperturbed graphene, but with
$t=(2t_1+t_2)/3$.
Diagonalizing Eq.~(\ref{hgap}),
we find Eq.~(\ref{gap}).
\end{widetext}


\end{document}